%% file: main.tex
\documentclass[11pt]{article}
\usepackage[utf8]{inputenc}
\usepackage[T1]{fontenc}
\usepackage{lmodern}
\usepackage[margin=1in]{geometry}
\usepackage{microtype}
\usepackage{booktabs}
\usepackage{longtable}
\usepackage{array}
\usepackage{graphicx}
\usepackage{float}
\usepackage{adjustbox}
\usepackage{amsmath,amssymb}
\usepackage{tikz}
\usetikzlibrary{arrows.meta,positioning,shapes.geometric,fit}
\usepackage{xcolor}
\usepackage{hyperref}
\usepackage{caption}
\usepackage{setspace}
\usepackage{pdflscape}
\usepackage{enumitem}
\usepackage{url}
\usepackage{needspace}
\hypersetup{colorlinks=true,linkcolor=blue!50!black,citecolor=blue!50!black,urlcolor=blue!50!black}
\graphicspath{{fig/}}
\newcommand{\orcidlink}{\href{https://orcid.org/0000-0002-4189-961X}{0000-0002-4189-961X}}
\newcommand{\zenododoi}{\href{https://doi.org/10.5281/zenodo.21134715}{10.5281/zenodo.21134715}}

\title{IMF Programs and Growth: A Source-Informed Robustness Reanalysis}
\author{Ricardo Alonzo Fern\'andez Salguero\\ORCID: \orcidlink}
\date{2026}

\begin{document}
\maketitle

\begin{abstract}
This article reassesses the meta-analytic evidence on the effect of International Monetary Fund programs on economic growth. The point of departure is the influential meta-analysis by Balima and Sokolova, which assembles 994 estimates from 36 studies and reports a positive average effect with substantial heterogeneity \cite{balima2021}. The reanalysis presented here imposes four stricter requirements. It treats the study, rather than the reported estimate, as the primary inferential unit; it uses a source-informed classification of causal credibility; it models within-study dependence through study aggregation, correlated-effects sensitivity, multilevel CR2 inference, and robust variance estimation; and it evaluates publication-bias sensitivity through Egger, PET, PEESE, WAAP-like top-precision analysis, p-curve diagnostics, trim-and-fill, and exploratory selection models \cite{egger1997,stanley2014,vevea1995,andrewskasy2019}. The central result is not that IMF programs reduce growth everywhere, nor that the true effect is exactly zero. The result is narrower and stronger: the positive average effect in the aggregate literature is not robust once dependence, publication selection, heterogeneity, and credibility of identification are treated as first-order concerns. In the most defensible specifications, the average effect is statistically indistinguishable from zero, while equivalence to a substantively negligible effect is only partially supported and depends on the chosen equivalence bound.
\end{abstract}

\noindent\textbf{How to cite:} Fern\'andez Salguero, Ricardo Alonzo. 2026. \textit{IMF Programs and Growth: A Source-Informed Robustness Reanalysis}. Zenodo. DOI: \zenododoi.


\section{Problem}

The empirical literature on IMF programs and economic growth has never converged on a single answer. Some studies estimate positive growth effects, especially when programs are evaluated after stabilization, among lower-income countries, or under longer horizons. Other studies find null or negative effects, especially when the selection of countries into IMF programs is modeled explicitly or when the contractionary component of adjustment is estimated during the program period. This disagreement is expected. Countries do not enter IMF programs randomly. They enter when balance-of-payments pressures, fiscal stress, reserve losses, debt rollover problems, or political constraints make external support valuable. The counterfactual path without a program is not observed, and therefore the estimand is not a raw difference between program and non-program observations. It is a causal contrast against an unobserved trajectory.

This selection problem is central in the older evaluation literature. Goldstein and Montiel identify the methodological pitfalls of multicountry program evaluation, Bird frames the policy problem of whether programs work and under what institutional conditions, and Dicks-Mireaux, Mecagni, and Schadler evaluate IMF lending to low-income countries with attention to non-random participation \cite{goldstein1986,bird2001,dicksmireaux2000}. Several influential studies then move toward explicit counterfactual or selection-aware designs. Przeworski and Vreeland estimate negative growth effects during program participation; Hardoy uses matching to reassess growth effects; Barro and Lee use political-economy instruments; Atoyan and Conway compare matching and instrumental-variable estimators; Dreher separates programs, loans, and compliance; Bas and Stone model adverse selection; Binder and Bluhm study conditional effects; Bal-Gunduz and Bird and Rowlands focus on low-income countries; and Newiak and Willems use synthetic-control evidence for non-financial programs \cite{przeworski2000,hardoy2003,barro2005,atoyan2006,dreher2006,bas2014,binder2017,balgunduz2016,bird2017,newiak2017}.

Balima and Sokolova \cite{balima2021} make a major contribution by collecting 994 estimates from 36 studies and by documenting extensive heterogeneity in the IMF-growth literature. Their meta-analysis is an important benchmark because it translates a fragmented literature into a common empirical object. The question addressed here is stricter. Does the positive average reported in the broader literature survive when estimates from the same study are no longer treated as independent, when the credibility of the source design is coded from the underlying papers, and when publication bias and precision selection are evaluated jointly with heterogeneity? The answer is no. Positive estimates are present, sometimes numerous, but the positive average is fragile.

The distinction between estimates and studies is fundamental. The 994 reported estimates are not 994 independent experiments. A single article may contribute many specifications built from the same country sample, outcome definition, program measure, covariate set, and authorial research design. Repeated estimates from the same study share unobserved choices and sampling errors. Treating them as independent observations overstates precision. The stricter design in this paper therefore asks what happens when the inferential unit is the study, or when the covariance among estimates within a study is modeled rather than ignored.

A second distinction concerns identification. A method label is not a validity certificate. An instrumental-variable estimate can be weak if the instrument affects growth through channels other than IMF participation. A matching design can remain biased if selection depends on unobservables. A difference-in-differences design is credible only if the comparison path is plausible. A synthetic-control design depends on pre-fit and placebo diagnostics. A generalized evaluation estimator depends on the stability of the policy reaction function. The reanalysis therefore does not simply rank studies by whether they say ``IV'', ``DID'', or ``PSM''. It uses a source-informed classification that assigns studies to credibility tiers while retaining a cautious interpretation of every tier.

The claim advanced here is deliberately limited. The paper does not show that IMF programs always fail, and it does not prove a precise zero effect. It shows that the existing meta-analytic evidence does not robustly establish a positive average causal effect of IMF programs on growth. In the source-informed run, effect-level random-effects specifications are often positive and statistically significant. However, the study-level estimates, multilevel CR2 estimates, RVE estimates, and precision-adjusted estimates are small and statistically indistinguishable from zero. This pattern is the empirical core of the paper.

\section{Identification}

The cleaned replication dataset contains 994 effect estimates from 36 studies. Each observation includes an effect estimate, a reported standard error, the study identifier, publication and design indicators, horizon variables, sample descriptors, and additional covariates from the original meta-analytic coding. The reanalysis adds a source-informed study classification based on the methods used in the primary studies. The classification is not a full risk-of-bias instrument. It is a structured way to avoid treating weak comparisons, partial counterfactual designs, and stronger quasi-causal designs as equivalent evidence.

The coding rule is transparent. Difference-in-differences, PSM-DID when identifiable, external or political instrumental variables with explicit selection adjustment, and synthetic-control evidence when isolated and diagnosed are treated as stronger quasi-causal designs, following the identification concerns raised in the source studies \cite{barro2005,atoyan2006,hardoy2003,newiak2017}. Propensity-score matching alone, generalized evaluation estimators, generic instrumental variables with unclear exclusion restrictions, conditional-effect panel models, and panel fixed-effects designs with partial controls are treated as medium partial counterfactual designs \cite{bas2014,binder2017,balgunduz2016,bird2017}. OLS, before-after comparisons, and designs that do not construct a credible counterfactual are treated as weak or without a valid counterfactual. Table \ref{tab:rubric} gives the codebook, Table \ref{tab:classes} reports the distribution of the analytical sample, and Table \ref{tab:methods} reports the method families.

\input{tbl/t01.tex}

\input{tbl/t02.tex}

\input{tbl/t03.tex}

The primary statistical problem is dependence. The conventional random-effects model treats the observed estimate $\hat\theta_{ij}$ from estimate $i$ in study $j$ as
\[
\hat\theta_{ij}=\theta_{ij}+e_{ij}, \qquad e_{ij}\sim N(0,s_{ij}^{2}),
\]
where $s_{ij}$ is the reported standard error. A simple random-effects model writes
\[
\theta_{ij}=\mu+u_{ij}, \qquad u_{ij}\sim N(0,\tau^{2}).
\]
That specification is useful descriptively but incomplete when a study contributes many related estimates. The correlated-effects sensitivity used here allows the sampling covariance between estimates from the same study to be nonzero:
\[
\operatorname{Cov}(e_{ij},e_{\ell j})=\rho s_{ij}s_{\ell j},\qquad i\neq \ell.
\]
The grid $\rho\in\{0,0.2,0.5,0.8\}$ is not an estimated truth. It is a sensitivity device. The value $\rho=0$ represents the conventional independence assumption, $\rho=0.2$ represents weak within-study dependence, $\rho=0.5$ represents a moderate working correlation that is plausible when specifications share samples and design choices, and $\rho=0.8$ represents a high-dependence stress test. If the positive effect only survives at low or zero correlation, it is not robust to dependence.

The implementation follows standard random-effects and multilevel meta-analysis machinery in \texttt{metafor}, robust variance estimation for dependent effects, and small-sample cluster-robust inference for meta-regression \cite{viechtbauer2010,hedges2010,pustejovsky2022}. Publication-bias diagnostics follow the funnel-asymmetry logic of Egger, PET and PEESE meta-regression approximations, weight-function and selection approaches, and identification-aware correction for publication selection \cite{egger1997,stanley2014,vevea1995,andrewskasy2019}.

The source-informed meta-regressions use the general form
\[
\hat\theta_{ij}=\alpha+X_{ij}\beta+\gamma s_{ij}+u_j+v_{ij}+e_{ij},
\]
where $X_{ij}$ includes causal class, method, horizon, sample, publication status, IMF-staff status, fixed-effect indicators, and design variables. The $s_{ij}$ term corresponds to PET-type publication-bias sensitivity; replacing $s_{ij}$ with $s_{ij}^{2}$ gives PEESE-type sensitivity. These regressions do not prove that one method causally changes the reported effect. They diagnose whether the literature's reported effects are systematically related to design, precision, and classification.

\begin{figure}[H]
\centering
\begin{tikzpicture}[node distance=1.1cm and 1.6cm, every node/.style={font=\small,align=center}]
\node[draw,rounded corners,fill=blue!8,minimum width=3.1cm,minimum height=0.85cm] (data) {994 estimates\\36 studies};
\node[draw,rounded corners,fill=blue!8,right=of data,minimum width=3.2cm,minimum height=0.85cm] (class) {Source-informed\\causal classes};
\node[draw,rounded corners,fill=blue!8,right=of class,minimum width=3.3cm,minimum height=0.85cm] (dep) {Dependence\\study, CR2, RVE};
\node[draw,rounded corners,fill=blue!8,below=of class,minimum width=3.6cm,minimum height=0.85cm] (pub) {Publication-bias\\PET, PEESE, WAAP};
\node[draw,rounded corners,fill=blue!8,right=of pub,minimum width=3.5cm,minimum height=0.85cm] (multi) {Specification\\curve and TOST};
\node[draw,rounded corners,fill=green!10,below=of dep,minimum width=4.1cm,minimum height=0.85cm] (concl) {Positive average not\\robust to stricter inference};
\draw[-{Latex},thick] (data) -- (class);
\draw[-{Latex},thick] (class) -- (dep);
\draw[-{Latex},thick] (class) -- (pub);
\draw[-{Latex},thick] (pub) -- (multi);
\draw[-{Latex},thick] (dep) -- (concl);
\draw[-{Latex},thick] (multi) -- (concl);
\end{tikzpicture}
\caption{Design of the source-informed analysis.}
\label{fig:designflow}
\end{figure}
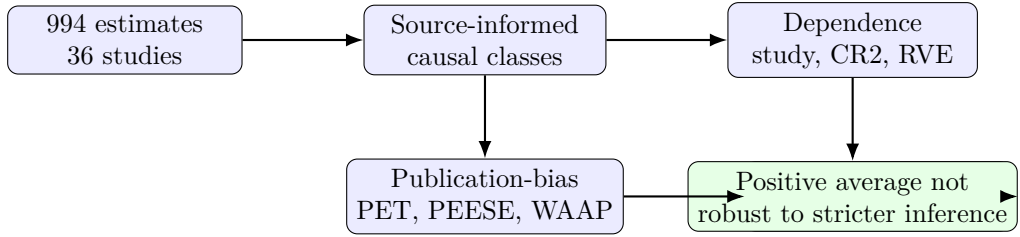

Outliers are handled through influence diagnostics, leave-one-study-out checks, study-level aggregation, robust-variance procedures, and mixture diagnostics rather than through mechanical deletion. This matters because extreme study means, such as very high positive averages in a small number of studies, may be substantively informative but should not dominate inference. The analysis reports the broad multiverse instead of selecting the most favorable or least favorable estimate.

\section{Evidence}

The first result is that effect-level meta-analysis reproduces the positive reading of the literature. Table \ref{tab:effectlevel} shows positive estimates in the full sample. The all-sample Paule-Mandel estimate is 0.257 with a confidence interval excluding zero. Yet the same table also shows why a single effect-level estimate is not enough. The estimates vary sharply across estimators and classes. In the stronger quasi-causal group, Paule-Mandel is positive, while REML is slightly negative. This estimator dependence is not a nuisance detail. It indicates that heterogeneity and weighting choices are doing substantial inferential work.

\input{tbl/t04.tex}

\begin{figure}[H]
\centering
\includegraphics[width=0.85\textwidth]{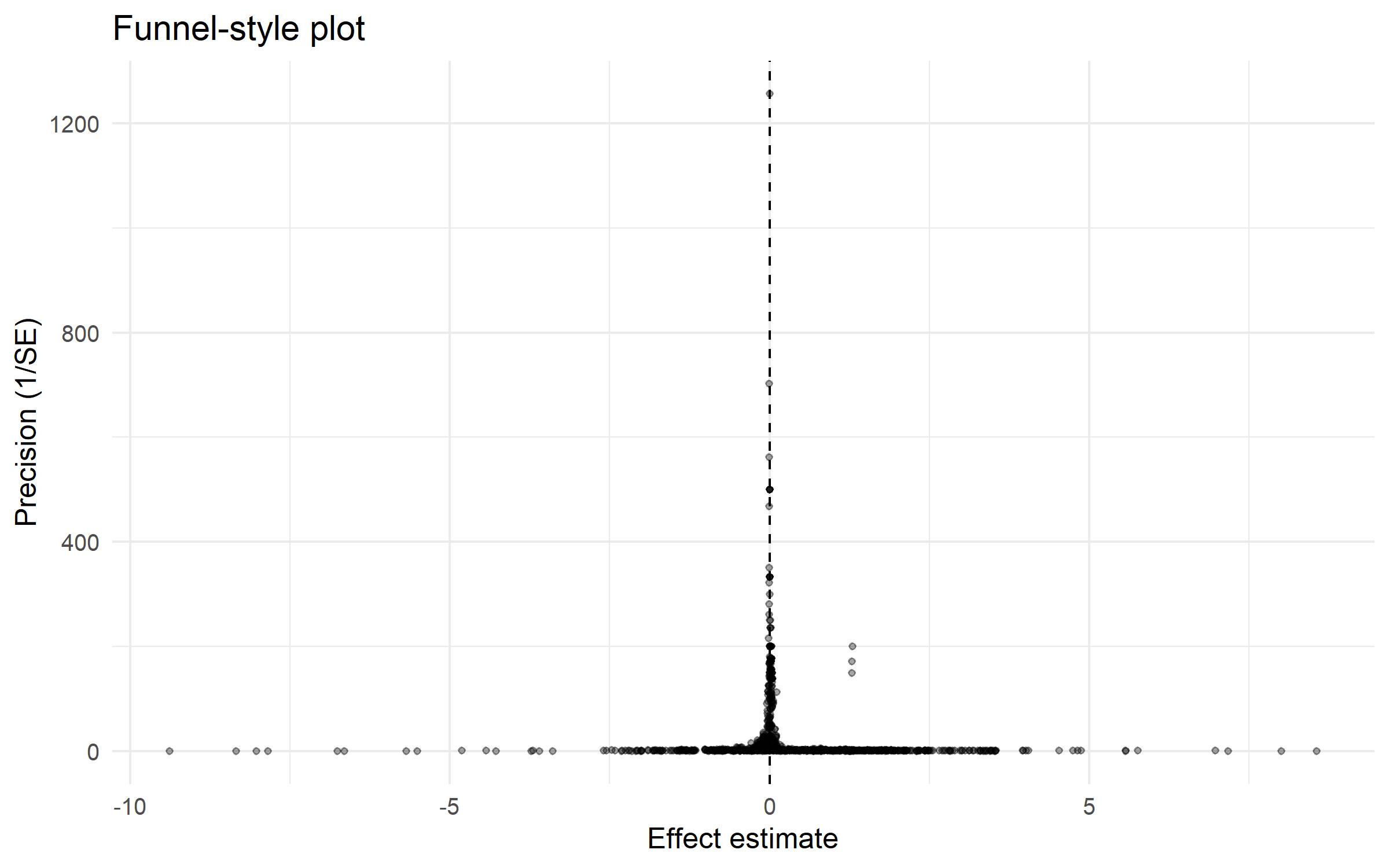}
\caption{Funnel-style plot of reported effects and precision in the full analytical sample.}
\label{fig:funnel}
\end{figure}

Figure \ref{fig:funnel} shows the funnel-style distribution of effects. It is not a clean symmetric cloud around a stable mean. The empirical distribution contains many small and imprecise estimates and a wide range of positive and negative reported effects. Publication-bias diagnostics are therefore necessary, but they cannot be interpreted apart from heterogeneity.

The most important result appears at the study level. Table \ref{tab:studyrho} reports the study-level random-effects results under alternative within-study correlations. In the full sample, the point estimate declines as $\rho$ rises: 0.163 at $\rho=0$, 0.095 at $\rho=0.5$, and 0.062 at $\rho=0.8$. None of these study-level estimates is statistically significant. In the stronger quasi-causal class, the estimates are small and close to zero under moderate and high correlation. The design filters that should be most persuasive do not deliver a stable positive study-level effect.

\input{tbl/t05.tex}

The multilevel CR2 and RVE results reinforce the study-level pattern. Table \ref{tab:cr2rve} shows that the multilevel CR2 intercept at $\rho=0.5$ is 0.036 with a confidence interval crossing zero. The RVE intercepts are about 0.043 and statistically insignificant across the working correlation grid. These estimates are among the most defensible in the package because they directly address dependent effect sizes without pretending that every reported estimate is an independent study.

\input{tbl/t06.tex}

Publication-bias diagnostics point in the same direction. Table \ref{tab:pubbias} reports PET and PEESE intercepts. In the full sample, the PET intercept is 0.017 and the PEESE intercept is 0.020; both are small and statistically insignificant. In the stronger quasi-causal class and in the strict DID/external-IV filter, the adjusted intercepts are slightly negative and statistically different from zero in the diagnostic regressions. These results should not be overread as evidence that IMF programs reduce growth. PET and PEESE can be unstable when heterogeneity is extreme. Their more defensible implication is that precision-adjusted estimates do not rescue the positive average.

\input{tbl/t07.tex}

The WAAP-like top-precision analysis is also unfavorable to the optimistic average. Table \ref{tab:waap} shows that the most precise quartile of estimates yields a small positive estimate in the full sample, but negative estimates in the stronger quasi-causal, medium partial counterfactual, strict DID/external-IV, and external-IV-only filters. Again, the conclusion is not a general negative causal effect. It is that the most precise estimates do not support a stable positive average.

\input{tbl/t08.tex}

The distinction between non-significance and equivalence is important. Table \ref{tab:tost} reports selected TOST results at the study level. The evidence does not generally prove equivalence under narrow bounds such as $\delta=0.10$. Equivalence appears only in some stronger-causal and higher-correlation settings under wider bounds. The correct interpretation is therefore precise: the effect is statistically indistinguishable from zero in the most defensible specifications, but exact or narrow substantive equivalence is not established uniformly.

\input{tbl/t09.tex}

\begin{figure}[H]
\centering
\includegraphics[width=0.94\textwidth]{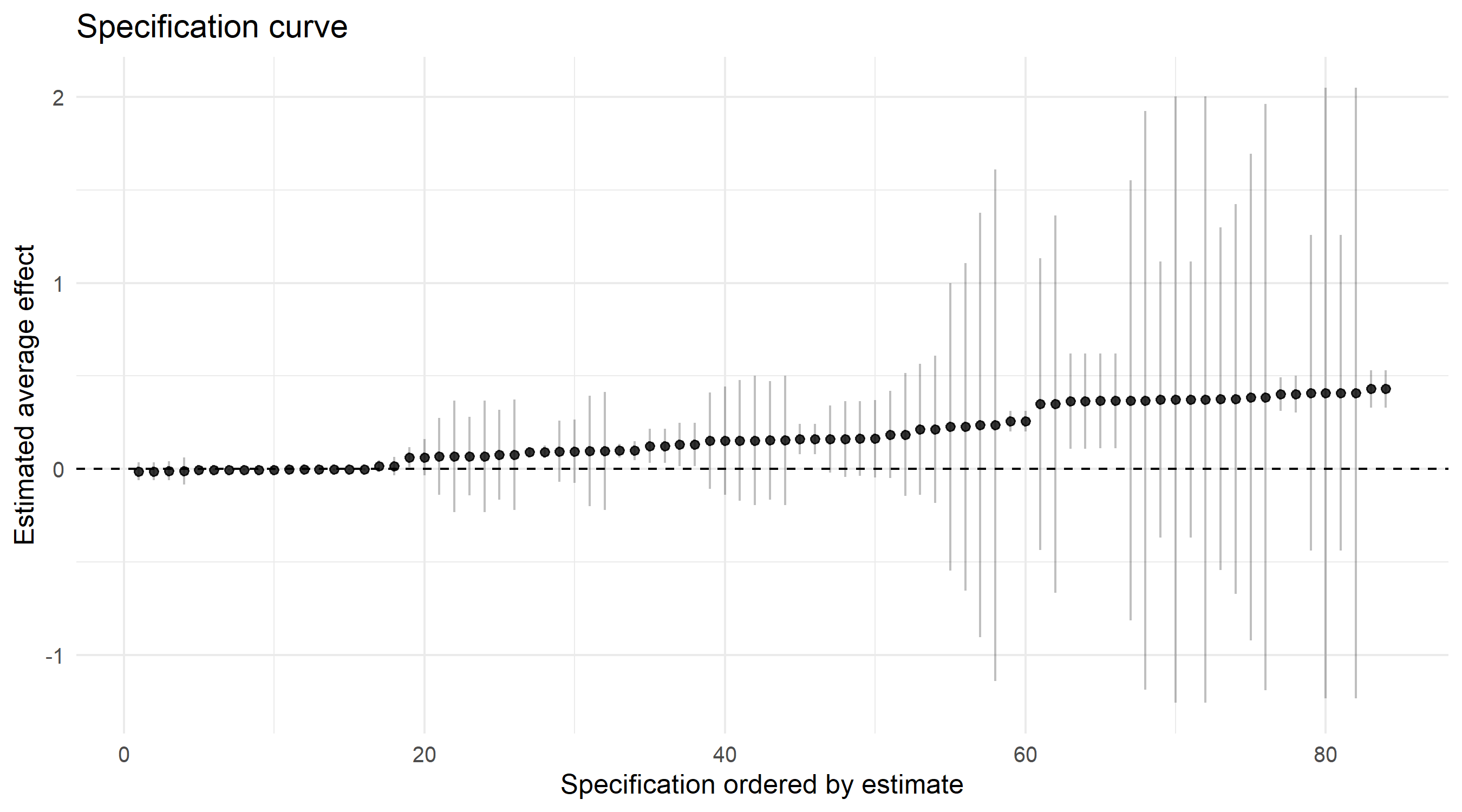}
\caption{Specification curve across aggregation levels, estimators, design filters, and dependence assumptions.}
\label{fig:spec}
\end{figure}

The specification curve summarizes the entire multiverse. Table \ref{tab:specsummary} shows that many specifications are positive, and some are significantly positive. Figure \ref{fig:spec} shows where those estimates sit. Positive significance is concentrated in effect-level, less conservative, or less dependence-adjusted specifications. Study-level, CR2, RVE, precision-adjusted, and stronger-causal specifications move toward zero or lose significance. The specification curve therefore supports a fragility interpretation rather than a null-by-assumption interpretation.

\input{tbl/t10.tex}

The p-curve and caliper diagnostics in Table \ref{tab:pcurve} show that the literature is not simply a mechanical product of bunching just below five percent significance. There are many strongly significant results. But that fact does not settle the causal question. A literature can contain real statistical signal and still fail to support a stable positive causal mean once dependence, design credibility, and publication selection are accounted for.

\input{tbl/t14.tex}

The sample-size diagnostic in Table \ref{tab:ndiag} explains why sample size should not be used as an instrument. Log sample size is weakly related to the standard error and is strongly explained by design covariates. This violates the logic of an instrument that shifts precision without directly changing the effect-generating process. Sample size captures country coverage, sample period, method, program definition, and data quality. It belongs in diagnostic and moderator analysis, not in an exclusion restriction.

\input{tbl/t11.tex}

The finite-mixture and exploratory selection results add one more layer. The mixture model describes the reported effects as a distribution with several components, not as draws from a single stable positive mean. The selection-model sensitivity is negative, but it should be interpreted cautiously because parametric selection models are fragile under dependence and high heterogeneity. Together, these diagnostics support the same conclusion: the literature is heterogeneous enough that the single positive pooled mean is not the most reliable summary.

\section{Interpretation}

The empirical pattern is not that every specification is zero. The empirical pattern is that the positive estimate depends on how the evidence is counted. If the unit is the reported estimate, and if many related estimates from the same paper are treated as independent, the pooled effect is positive. If the unit is the study, or if within-study dependence is modeled, the effect shrinks and loses significance. If source-informed causal credibility is imposed, the strongest designs do not produce a stable positive study-level estimate. If precision-adjusted publication-bias diagnostics are applied, the positive intercept becomes small or disappears.

This interpretation is consistent with the economics of IMF programs. A program can improve external liquidity, prevent disorderly default, coordinate expectations, and anchor macroeconomic policy \cite{bird2001,dicksmireaux2000}. It can also impose fiscal contraction, monetary tightening, exchange-rate adjustment, and structural reforms that depress output in the short run \cite{przeworski2000,dreher2006}. The sign of the estimated growth effect should therefore depend on timing, crisis severity, program type, financing terms, conditionality, political feasibility, and the counterfactual. A short-run stabilization effect is not the same estimand as a long-run institutional or credibility effect. This is why the average treatment effect across all designs and horizons has limited substantive meaning.

The credibility classification helps interpret the pattern. PSM-only designs can be informative, but they identify effects under selection on observables. If unobserved political capacity, reserve adequacy, debt structure, or reform commitment jointly affects IMF participation and subsequent growth, PSM can remain biased. IV designs are attractive because they target endogeneity, but political or geopolitical instruments are vulnerable if they affect growth through trade, aid, diplomatic alignment, financing access, or conflict channels. DID requires credible pretrends and treatment timing; in this dataset, the DID-only evidence comes from too few studies to carry a general conclusion. GEE and strategic selection approaches are valuable but depend on stability assumptions. The aggregate literature cannot be interpreted as if all these designs identify the same clean causal estimand.

The apparent dilution of the positive effect also has a reporting interpretation. Studies that report many specifications can have disproportionate influence at the effect level. If a paper with many estimates has a positive mean, it can dominate an analysis that treats every estimate equally. Study-level inference reduces this imbalance. This is not an arbitrary conservative choice. It is a adjustment for the fact that a paper is a research design, not merely a collection of independent numbers.

Policy interpretation should therefore be modest. The evidence does not justify a broad claim that IMF programs raise growth on average. It also does not justify a broad claim that IMF programs reduce growth in all contexts. The evidence is compatible with a distribution of effects: some positive, some null, some negative. The next substantive research question is not whether the IMF works in the abstract. It is when program design, country conditions, financing constraints, and implementation capacity make growth effects more likely to be positive or negative.

\begin{figure}[H]
\centering
\begin{tikzpicture}[x=1cm,y=1cm, every node/.style={font=\small,align=center}]
\draw[->,thick] (0,0) -- (12.7,0) node[right]{Stricter inference};
\draw[->,thick] (0,-0.4) -- (0,3.3) node[above]{Estimated average};
\draw[dashed,gray] (0,1.05) -- (12.2,1.05) node[right]{zero neighborhood};
\node[draw,rounded corners,fill=blue!10,minimum width=2.5cm,minimum height=0.8cm] at (1.6,2.65) {Effect-level PM\\0.257};
\node[draw,rounded corners,fill=blue!8,minimum width=2.8cm,minimum height=0.8cm] at (4.5,1.95) {Study-level\\$\rho=0.5$: 0.095};
\node[draw,rounded corners,fill=blue!6,minimum width=3.1cm,minimum height=0.8cm] at (7.7,1.35) {Multilevel CR2\\$\rho=0.5$: 0.036};
\node[draw,rounded corners,fill=blue!4,minimum width=3.1cm,minimum height=0.8cm] at (10.8,1.05) {Stronger causal\\near zero};
\draw[-{Latex},thick] (2.75,2.52) -- (3.2,2.05);
\draw[-{Latex},thick] (5.85,1.85) -- (6.15,1.42);
\draw[-{Latex},thick] (9.15,1.30) -- (9.25,1.08);
\end{tikzpicture}
\caption{The positive average weakens as dependence, study-level inference, and source-informed causal credibility are imposed.}
\label{fig:dilution}
\end{figure}
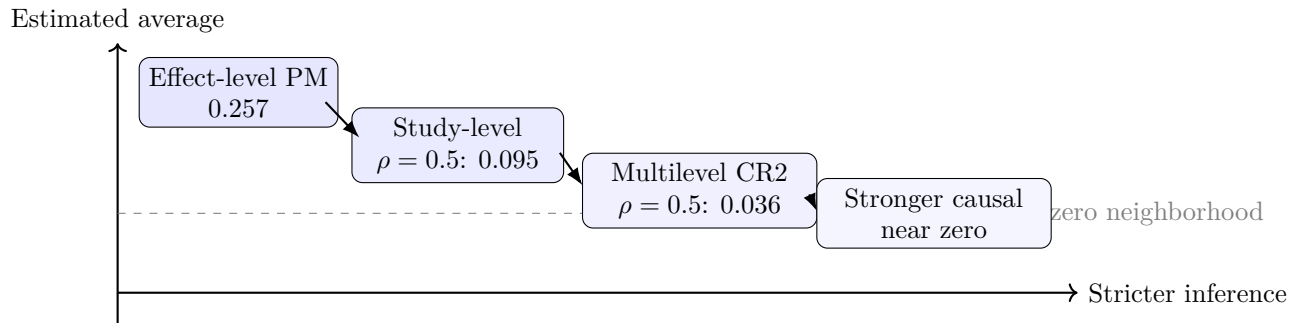

\section{Limits}

The most important limitation is power. Some high-credibility filters contain few studies. DID-only evidence is too thin for a strong conclusion, and external-IV-only evidence is too heterogeneous for a precise estimate. Therefore, failure to reject zero in these subgroups should not be misread as proof of no effect. It is evidence that the current literature does not robustly establish a positive effect under stricter inferential rules.

A second limitation is unresolved heterogeneity. The high $I^{2}$ values show that the studies are not estimating a single common parameter. The reanalysis addresses this through study-level aggregation, correlated-effects sensitivity, design filters, CR2, RVE, specification curves, and publication-bias diagnostics. These are substantial improvements, but they do not fully explain heterogeneity. A more complete design would code program type, financing size, conditionality intensity, debt distress, exchange-rate regime, political institutions, pre-program crisis severity, and post-program horizon. The cleaned data contain some horizon and sample indicators, but not enough detail to estimate all mechanisms credibly.

A third limitation is classification uncertainty. The source-informed coding improves on mechanical method dummies, but it is still a structured judgment. A journal submission should include double coding by independent reviewers or a full risk-of-bias appendix. The present package provides the manual audit and study-level classification so that readers can revise the coding and rerun the analysis.

A fourth limitation concerns publication-bias adjustments. Egger, PET, PEESE, WAAP-like, trim-and-fill, and selection models each answer different questions and each can fail under high heterogeneity. They are therefore interpreted as triangulation. Their common message is that precision adjustment does not support a stable positive average; their individual point estimates should not be treated as definitive.

A fifth limitation concerns equivalence. The most defensible estimates are statistically indistinguishable from zero, but equivalence to a substantively negligible bound is not uniformly proven. The conclusion should therefore be written as a robustness claim: the positive average is not robust. It should not be written as a proof that the true effect is exactly zero.

\input{tbl/t12.tex}

\section{Replication}

The replication package contains the cleaned Stata dataset, the source-informed study classification, the manual source-paper audit, the R script, generated tables, generated figures, model objects, logs, and this manuscript. The script expects a data directory containing \path{data.dta}, \path{class.csv}, and \path{audit.csv}. It can be run with:
\begin{quote}
\ttfamily\footnotesize
Rscript scr/run.R --data\_dir=data --out=out
\end{quote}
The run should report complete source-informed coverage. The package includes the output tables and session information so that the numerical results can be checked without rerunning the full pipeline.

\input{tbl/t13.tex}

The preferred replication sequence is simple. First, verify that all 36 studies receive a source-informed classification. Second, reproduce the effect-level random-effects models. Third, reproduce the study-level results over the $\rho$ grid. Fourth, inspect the multilevel CR2 and RVE intercepts. Fifth, examine PET, PEESE, WAAP-like, p-curve, and selection-model sensitivity as diagnostics rather than as a single decisive adjustment. Sixth, compare the specification curve to the main text to verify that the positive result is concentrated in less conservative specifications.

\section{References}
\begingroup
\renewcommand{\refname}{}

\endgroup

\end{document}

%% file: tbl/t01.tex
\begin{table}[H]
\centering
\caption{Design codebook.}
\label{tab:rubric}
\small
\begin{adjustbox}{max width=\textwidth}
\begin{tabular}{p{0.23\textwidth}p{0.35\textwidth}p{0.35\textwidth}}
\toprule
Class & Included designs & Identification caveat \\
\midrule
Stronger quasi-causal & Difference-in-differences; PSM-DID where identifiable; external or political IV with explicit selection adjustment; synthetic-control evidence when isolated and diagnosed. & Requires pretrend, exclusion, first-stage, placebo, or pre-fit checks; high causal ambition but still audited as fallible. \\
Medium partial counterfactual & PSM-only; generalized evaluation estimators; generic IV with unclear exclusion; panel FE with substantial controls; conditional-effect models. & Improves over raw comparisons but remains vulnerable to unobservables, unstable policy-reaction functions, weak instruments, or misspecified selection. \\
Weak or no valid counterfactual & OLS or FE without selection adjustment; before-after comparisons; designs that mainly compare program observations with observed non-program observations. & Useful descriptively, weak for causal inference under adverse selection into IMF programs. \\
\bottomrule
\end{tabular}
\end{adjustbox}
\end{table}

%% file: tbl/t02.tex
\begin{table}[H]
\centering
\caption{Causal classes.}
\label{tab:classes}
\small
\begin{adjustbox}{max width=\textwidth}
\begin{tabular}{lrrrrrr}
\toprule
Class & Effects & Studies & Mean & Median & Share > 0 & Share p < .05 \\
\midrule
Medium Partial Counterfactual & 446 & 19 & 0.450 & 0.355 & 0.637 & 0.294 \\
Stronger Quasi Causal & 251 & 5 & 0.310 & 0.007 & 0.554 & 0.199 \\
Weak No Valid Counterfactual & 297 & 12 & 0.262 & 0.028 & 0.761 & 0.505 \\
\bottomrule
\end{tabular}
\end{adjustbox}
\end{table}

%% file: tbl/t03.tex
\begin{table}[H]
\centering
\caption{Method families.}
\label{tab:methods}
\small
\begin{adjustbox}{max width=\textwidth}
\begin{tabular}{lrrrrrrr}
\toprule
Method & Effects & Studies & Mean & Median & Mean SE & Share > 0 & Share p < .05 \\
\midrule
OLS/FE/other & 380 & 15 & 0.114 & 0.024 & 0.527 & 0.666 & 0.461 \\
PSM & 285 & 6 & 0.886 & 0.970 & 0.993 & 0.796 & 0.326 \\
IV & 165 & 7 & 0.107 & 0.000 & 0.376 & 0.491 & 0.200 \\
DID & 111 & 1 & 0.494 & 0.357 & 1.169 & 0.622 & 0.198 \\
GEE & 50 & 6 & -0.211 & -0.197 & 0.784 & 0.360 & 0.140 \\
BA & 3 & 1 & -0.490 & -0.220 & 0.600 & 0.333 & 0.333 \\
\bottomrule
\end{tabular}
\end{adjustbox}
\end{table}

%% file: tbl/t04.tex
\begin{table}[H]
\centering
\caption{Effect-level estimates.}
\label{tab:effectlevel}
\small
\begin{adjustbox}{max width=\textwidth}
\begin{tabular}{lrrrrrrrrr}
\toprule
Sample & Effects & Studies & Estimate & CI low & CI high & p & tau2 & I2 & Estimator \\
\midrule
All / PM & 994 & 36 & 0.257 & 0.200 & 0.313 & 5.33e-19 & 0.506 & 100.0 & PM \\
All / REML & 994 & 36 & 0.091 & 0.067 & 0.115 & 8.62e-14 & 0.056 & 99.8 & REML \\
stronger quasi causal / PM & 251 & 5 & 0.124 & 0.032 & 0.216 & 0.008 & 0.316 & 100.0 & PM \\
stronger quasi causal / REML & 251 & 5 & -0.005 & -0.007 & -0.002 & 4.16e-04 & 2.90e-05 & 25.9 & REML \\
medium partial counterfactual / PM & 446 & 19 & 0.430 & 0.329 & 0.531 & 5.81e-17 & 0.668 & 99.6 & PM \\
medium partial counterfactual / REML & 446 & 19 & 0.402 & 0.312 & 0.492 & 2.51e-18 & 0.479 & 99.5 & REML \\
weak no valid counterfactual / PM & 297 & 12 & 0.160 & 0.078 & 0.241 & 1.25e-04 & 0.368 & 100.0 & PM \\
weak no valid counterfactual / REML & 297 & 12 & 0.098 & 0.062 & 0.134 & 1.07e-07 & 0.058 & 99.9 & REML \\
\bottomrule
\end{tabular}
\end{adjustbox}
\end{table}

%% file: tbl/t05.tex
\begin{table}[H]
\centering
\caption{Study-level estimates.}
\label{tab:studyrho}
\small
\begin{adjustbox}{max width=\textwidth}
\begin{tabular}{lrrrrrrrr}
\toprule
Sample & rho & Studies & Estimate & CI low & CI high & p & tau2 & I2 \\
\midrule
All & 0.0 & 36 & 0.163 & -0.046 & 0.371 & 0.122 & 0.328 & 100.0 \\
Stronger Quasi Causal & 0.0 & 5 & 0.068 & -0.231 & 0.367 & 0.563 & 0.057 & 100.0 \\
Medium Partial Counterfactual & 0.0 & 19 & 0.153 & -0.195 & 0.501 & 0.367 & 0.440 & 99.9 \\
Weak No Valid Counterfactual & 0.0 & 12 & 0.214 & -0.182 & 0.609 & 0.259 & 0.340 & 100.0 \\
Strict Did Or External Iv & 0.0 & 11 & 0.377 & -0.671 & 1.424 & 0.441 & 2.303 & 100.0 \\
Iv External Only & 0.0 & 8 & 0.386 & -1.191 & 1.963 & 0.581 & 3.416 & 100.0 \\
All & 0.5 & 36 & 0.095 & -0.075 & 0.264 & 0.264 & 0.161 & 100.0 \\
Stronger Quasi Causal & 0.5 & 5 & -0.011 & -0.083 & 0.062 & 0.701 & 0.002 & 99.0 \\
Medium Partial Counterfactual & 0.5 & 19 & 0.096 & -0.221 & 0.413 & 0.532 & 0.281 & 99.3 \\
Weak No Valid Counterfactual & 0.5 & 12 & 0.153 & -0.138 & 0.444 & 0.272 & 0.138 & 99.9 \\
Strict Did Or External Iv & 0.5 & 11 & 0.226 & -0.653 & 1.105 & 0.579 & 1.422 & 100.0 \\
Iv External Only & 0.5 & 8 & 0.236 & -1.140 & 1.611 & 0.697 & 2.466 & 100.0 \\
All & 0.8 & 36 & 0.062 & -0.082 & 0.206 & 0.391 & 0.098 & 99.9 \\
Stronger Quasi Causal & 0.8 & 5 & -0.001 & -0.019 & 0.017 & 0.889 & 1.07e-04 & 74.9 \\
Medium Partial Counterfactual & 0.8 & 19 & 0.065 & -0.231 & 0.361 & 0.650 & 0.205 & 98.6 \\
Weak No Valid Counterfactual & 0.8 & 12 & 0.126 & -0.115 & 0.366 & 0.274 & 0.081 & 99.8 \\
Strict Did Or External Iv & 0.8 & 11 & 0.142 & -0.618 & 0.902 & 0.686 & 0.954 & 100.0 \\
Iv External Only & 0.8 & 8 & 0.154 & -1.082 & 1.390 & 0.777 & 1.922 & 100.0 \\
\bottomrule
\end{tabular}
\end{adjustbox}
\end{table}

%% file: tbl/t06.tex
\begin{table}[H]
\centering
\caption{CR2 and RVE estimates.}
\label{tab:cr2rve}
\small
\begin{adjustbox}{max width=\textwidth}
\begin{tabular}{lrrrrrr}
\toprule
Estimator & rho & Estimate & SE & CI low & CI high & p \\
\midrule
Multilevel CR2 & 0.0 & 0.172 & 0.096 & -0.024 & 0.367 & 0.083 \\
Multilevel CR2 & 0.2 & 0.066 & 0.060 & -0.057 & 0.189 & 0.280 \\
Multilevel CR2 & 0.5 & 0.036 & 0.059 & -0.085 & 0.158 & 0.546 \\
Multilevel CR2 & 0.8 & 0.106 & 0.079 & -0.056 & 0.269 & 0.190 \\
RVE robumeta & 0.2 & 0.042 & 0.070 & -0.112 & 0.197 & 0.559 \\
RVE robumeta & 0.5 & 0.043 & 0.071 & -0.112 & 0.197 & 0.558 \\
RVE robumeta & 0.8 & 0.043 & 0.071 & -0.112 & 0.197 & 0.557 \\
\bottomrule
\end{tabular}
\end{adjustbox}
\end{table}

%% file: tbl/t07.tex
\begin{table}[H]
\centering
\caption{PET and PEESE.}
\label{tab:pubbias}
\small
\begin{adjustbox}{max width=\textwidth}
\begin{tabular}{lrrrr}
\toprule
Sample & Model & Intercept & SE & p \\
\midrule
All & PET intercept & 0.017 & 0.017 & 0.333 \\
All & PEESE intercept & 0.020 & 0.019 & 0.284 \\
Stronger Quasi Causal & PET intercept & -0.003 & 0.001 & 5.94e-05 \\
Stronger Quasi Causal & PEESE intercept & -0.003 & 0.001 & 0.003 \\
Medium Partial Counterfactual & PET intercept & -0.011 & 0.009 & 0.259 \\
Medium Partial Counterfactual & PEESE intercept & -0.005 & 0.006 & 0.440 \\
Weak No Valid Counterfactual & PET intercept & 0.037 & 0.027 & 0.169 \\
Weak No Valid Counterfactual & PEESE intercept & 0.044 & 0.028 & 0.109 \\
Strict Did Or External Iv & PET intercept & -0.003 & 0.001 & 2.15e-05 \\
Strict Did Or External Iv & PEESE intercept & -0.003 & 0.001 & 0.001 \\
Iv External Only & PET intercept & -0.003 & 0.001 & 0.001 \\
Iv External Only & PEESE intercept & -0.003 & 0.001 & 0.001 \\
\bottomrule
\end{tabular}
\end{adjustbox}
\end{table}

%% file: tbl/t08.tex
\begin{table}[H]
\centering
\caption{Top-precision estimates.}
\label{tab:waap}
\small
\begin{adjustbox}{max width=\textwidth}
\begin{tabular}{lrrrrrrrr}
\toprule
Sample & Effects & Studies & Estimate & CI low & CI high & p & tau2 & I2 \\
\midrule
All & 249 & 12 & 0.021 & 0.002 & 0.039 & 0.030 & 0.022 & 99.8 \\
Stronger Quasi Causal & 64 & 4 & -0.005 & -0.008 & -0.001 & 0.008 & 8.18e-05 & 79.6 \\
Medium Partial Counterfactual & 112 & 12 & -0.075 & -0.133 & -0.017 & 0.011 & 0.058 & 98.9 \\
Weak No Valid Counterfactual & 78 & 3 & 0.070 & 0.015 & 0.125 & 0.014 & 0.060 & 100.0 \\
Strict Did Or External Iv & 75 & 4 & -0.005 & -0.008 & -0.002 & 0.001 & 6.53e-05 & 72.7 \\
Iv External Only & 39 & 4 & -0.005 & -0.008 & -0.002 & 0.001 & 1.03e-04 & 89.0 \\
Did Only & 37 & 3 & 0.401 & -0.087 & 0.889 & 0.104 & 1.492 & 70.8 \\
\bottomrule
\end{tabular}
\end{adjustbox}
\end{table}

%% file: tbl/t09.tex
\begin{table}[H]
\centering
\caption{Equivalence tests.}
\label{tab:tost}
\small
\begin{adjustbox}{max width=\textwidth}
\begin{tabular}{lrrrrr}
\toprule
Sample & rho & delta & Estimate & TOST p max & Equivalent \\
\midrule
All & 0.5 & 0.10 & 0.095 & 0.475 & No \\
All & 0.5 & 0.20 & 0.095 & 0.108 & No \\
All & 0.5 & 0.30 & 0.095 & 0.009 & Yes \\
Stronger Quasi Causal & 0.5 & 0.10 & -0.011 & 0.013 & Yes \\
Stronger Quasi Causal & 0.5 & 0.20 & -0.011 & 0.001 & Yes \\
Stronger Quasi Causal & 0.5 & 0.30 & -0.011 & 1.89e-04 & Yes \\
All & 0.8 & 0.10 & 0.062 & 0.296 & No \\
All & 0.8 & 0.20 & 0.062 & 0.030 & Yes \\
All & 0.8 & 0.30 & 0.062 & 0.001 & Yes \\
Stronger Quasi Causal & 0.8 & 0.10 & -0.001 & 5.36e-05 & Yes \\
Stronger Quasi Causal & 0.8 & 0.20 & -0.001 & 3.35e-06 & Yes \\
Stronger Quasi Causal & 0.8 & 0.30 & -0.001 & 6.61e-07 & Yes \\
\bottomrule
\end{tabular}
\end{adjustbox}
\end{table}

%% file: tbl/t10.tex
\begin{table}[H]
\centering
\caption{Specification curve.}
\label{tab:specsummary}
\small
\begin{adjustbox}{max width=\textwidth}
\begin{tabular}{lrrrrrrrr}
\toprule
Specs & Positive & Negative & p<.05 & Positive p<.05 & Negative p<.05 & Median & Min & Max \\
\midrule
84.0 & 68.0 & 16.0 & 29.0 & 21.0 & 8.0 & 0.153 & -0.013 & 0.430 \\
\bottomrule
\end{tabular}
\end{adjustbox}
\end{table}

%% file: tbl/t14.tex
\begin{table}[H]
\centering
\caption{P-curve.}
\label{tab:pcurve}
\small
\begin{adjustbox}{max width=\textwidth}
\begin{tabular}{lrrrr}
\toprule
Just below .05 & Just above .05 & p in [.025,.05) & p < .025 & Total \\
\midrule
30 & 25 & 81 & 250 & 994 \\
\bottomrule
\end{tabular}
\end{adjustbox}
\end{table}

%% file: tbl/t11.tex
\begin{table}[H]
\centering
\caption{Sample size.}
\label{tab:ndiag}
\small
\begin{adjustbox}{max width=\textwidth}
\begin{tabular}{lrrrrrr}
\toprule
Diagnostic & N & Coef. log(N) & Cluster SE & F or R2 & p & R2 \\
\midrule
effect\_se on log sample size & 994 & 0.019 & 0.108 & 0.030 & 0.864 & 4.54e-04 \\
se2 on log sample size & 994 & -0.436 & 0.496 & 0.771 & 0.380 & 0.001 \\
log sample size explained by design covariates & 994 &  &  & 0.700 & 7.35e-241 & 0.700 \\
\bottomrule
\end{tabular}
\end{adjustbox}
\vspace{0.25em}
\begin{minipage}{0.96\textwidth}\footnotesize Sample size is not used as an instrument. The diagnostic shows weak first-stage behavior for precision and strong confounding with study design.\end{minipage}
\end{table}

%% file: tbl/t12.tex
\begin{landscape}
\begin{longtable}{p{0.18\linewidth}p{0.35\linewidth}p{0.39\linewidth}}
\caption{Source audit.}\label{tab:sourceaudit}\\
\toprule
Source & Method from sources & Credibility note \\
\midrule
\endfirsthead
\toprule
Source & Method from sources & Credibility note \\
\midrule
\endhead
Goldstein \& Montiel (1986) & Generalized evaluation estimator foundation / multicountry evaluation pitfalls & Historically important, but assumes stable policy reaction/counterfactual; not modern causal design. \\
Hardoy (2003) & Matching and difference-in-differences matching & Better than simple PSM if DiD-matching is used; still depends on observables/common support and parallel trends. \\
Atoyan \& Conway (2006) & Censored-sample, IV, and matching comparison & Important estimator comparison; results differ by estimator, so should not be collapsed into one generic quality tier. \\
Barro \& Lee (2005) & IV selection using political/proximity type instruments & Causal ambition high; exclusion restriction controversial because geopolitics may affect growth and IMF access. \\
Dreher (2006) & Endogeneity-adjusted program/loan/compliance models & High relevance because it explicitly accounts for endogeneity and finds programs reduce growth; IV validity still needs audit. \\
Bas \& Stone (2014) & Strategic selection model for adverse selection & Should be a separate source-informed class, not OLS/other; important for selection mechanisms. \\
Binder \& Bluhm (2017) & State-dependent panel model; conditional effects and selection & Quality depends on conditional-effect specification; not equivalent to simple matching or OLS. \\
Bal-Gunduz (2016) & Propensity score matching in LIC homogeneous sample & Better PSM due to homogeneous financing events and richer selection model, but still observables-only. \\
Mumssen et al. (2013) & LIC short/longer-term impact; PSM; separates longer-term engagement and short-term financing & Useful because it separates program types/horizons, but IMF staff status should be modeled. \\
Bird \& Rowlands (2017) & LIC-specific participation model with PSM & PSM-only; positive effects up to two years; credible for observables but not unobservable selection. \\
Newiak \& Willems (2017) & Synthetic control for non-financial PSI programs & Should be isolated as SCM, not PSM; small number of cases and diagnostics weaken certainty. \\
Ozturk (2008/2011) & GEE applications & Not weak OLS, but relies on stable policy reaction function. \\
\bottomrule
\end{longtable}
\end{landscape}

%% file: tbl/t13.tex
\begin{landscape}
\begin{longtable}{rrrlrrrr}
\caption{Study classes.}\label{tab:studyclass}\\
\toprule
ID & Effects & Year & Method & Causal class & Mean & Median & Share > 0 \\
\midrule
\endfirsthead
\toprule
ID & Effects & Year & Method & Causal class & Mean & Median & Share > 0 \\
\midrule
\endhead
1 & 5 & 2006 & PSM & medium partial counterfactual & 0.017 & -0.029 & 0.400 \\
2 & 12 & 2003 & OLS\_FE\_OTHER & medium partial counterfactual & 1.242 & 1.035 & 0.750 \\
3 & 24 & 2005 & IV & stronger quasi causal & -0.079 & -0.024 & 0.042 \\
4 & 14 & 2014 & OLS\_FE\_OTHER & weak no valid counterfactual & 0.049 & 0.057 & 0.857 \\
5 & 42 & 2017 & IV & stronger quasi causal & 0.537 & 0.012 & 0.810 \\
6 & 57 & 2017 & PSM & medium partial counterfactual & 0.825 & 1.110 & 0.789 \\
7 & 25 & 2000 & OLS\_FE\_OTHER & weak no valid counterfactual & -0.185 & -0.979 & 0.440 \\
8 & 40 & 2005 & OLS\_FE\_OTHER & medium partial counterfactual & -0.054 & -0.075 & 0.300 \\
9 & 4 & 1994 & IV & medium partial counterfactual & 0.970 & 0.945 & 1.000 \\
10 & 1 & 2000 & GEE & medium partial counterfactual & 1.374 & 1.374 & 1.000 \\
11 & 24 & 2006 & OLS\_FE\_OTHER & medium partial counterfactual & -1.992 & -0.530 & 0.125 \\
12 & 10 & 2006 & OLS\_FE\_OTHER & medium partial counterfactual & -0.092 & -0.012 & 0.400 \\
13 & 8 & 2015 & OLS\_FE\_OTHER & weak no valid counterfactual & 2.821 & 2.136 & 1.000 \\
14 & 3 & 1986 & BA & weak no valid counterfactual & -0.490 & -0.220 & 0.333 \\
15 & 74 & 2016 & PSM & medium partial counterfactual & 1.372 & 1.356 & 0.986 \\
16 & 10 & 1987 & OLS\_FE\_OTHER & weak no valid counterfactual & 0.370 & 0.350 & 0.800 \\
17 & 96 & 2003 & PSM & medium partial counterfactual & 0.014 & 0.135 & 0.583 \\
18 & 19 & 2003 & IV & medium partial counterfactual & -0.339 & -0.730 & 0.368 \\
19 & 16 & 2003 & GEE & medium partial counterfactual & -0.608 & -0.699 & 0.125 \\
20 & 25 & 2004 & GEE & medium partial counterfactual & 0.086 & 0.090 & 0.560 \\
21 & 14 & 2012 & OLS\_FE\_OTHER & weak no valid counterfactual & 1.135 & 0.810 & 0.929 \\
22 & 20 & 2002 & IV & stronger quasi causal & -0.137 & -0.180 & 0.300 \\
23 & 6 & 1990 & GEE & medium partial counterfactual & -0.312 & -0.270 & 0.167 \\
24 & 6 & 2011 & OLS\_FE\_OTHER & weak no valid counterfactual & -1.641 & -0.713 & 0.167 \\
25 & 2 & 2004 & OLS\_FE\_OTHER & weak no valid counterfactual & 0.075 & 0.075 & 1.000 \\
26 & 51 & 2013 & PSM & medium partial counterfactual & 2.038 & 1.760 & 1.000 \\
27 & 2 & 2017 & PSM & medium partial counterfactual & -0.737 & -0.737 & 0.000 \\
28 & 111 & 2016 & DID & stronger quasi causal & 0.494 & 0.357 & 0.622 \\
29 & 12 & 2005 & OLS\_FE\_OTHER & weak no valid counterfactual & 1.242 & 1.035 & 0.750 \\
30 & 80 & 2013 & OLS\_FE\_OTHER & weak no valid counterfactual & 0.056 & 0.004 & 0.537 \\
31 & 2 & 1993 & IV & medium partial counterfactual & -1.329 & -1.329 & 0.000 \\
33 & 3 & 2004 & OLS\_FE\_OTHER & weak no valid counterfactual & 3.066 & 2.814 & 1.000 \\
35 & 1 & 2008 & GEE & medium partial counterfactual & -0.474 & -0.474 & 0.000 \\
36 & 1 & 2011 & GEE & medium partial counterfactual & -2.019 & -2.019 & 0.000 \\
37 & 54 & 2008 & IV & stronger quasi causal & 0.093 & 0.001 & 0.537 \\
38 & 120 & 2013 & OLS\_FE\_OTHER & weak no valid counterfactual & 0.185 & 0.026 & 0.958 \\
\bottomrule
\end{longtable}
\end{landscape}